\documentclass[a4paper]{jpconf}

\usepackage{enumitem}
\usepackage{graphicx}
\usepackage{color}

\usepackage[pdftex,colorlinks=true,bookmarks=false,citecolor=blue,urlcolor=blue]{hyperref} 

\begin{document}

\title{Status of GEO\,600}

\author{K~L~Dooley \footnote{Current address: The University of Mississippi, University, MS 38677, USA} for the LIGO Scientific Collaboration}
\address{Max-Planck-Institut f\"ur Gravitationsphysik
(Albert-Einstein-Institut) und Leibniz Universit\"at Hannover, Callinstr.\ 38,
D-30167 Hannover, Germany}

\ead{kate.dooley@aei.mpg.de}

\date{\today}

\begin{abstract}
The German-British laser-interferometric gravitational wave detector
GEO\,600 is in its 13th year of operation since its first lock in
2001. After participating in science runs with other first generation
detectors, GEO\,600 has continued collecting data as an astrowatch
instrument with a duty cycle of 62\% during the time when the other
detectors have gone offline to undergo substantial upgrades. Less
invasive upgrades to demonstrate advanced technologies and improve the
GEO\,600 sensitivity at high frequencies as part of the GEO-HF program
have additionally been carried out in parallel to data taking. We
report briefly on the status of GEO\,600.
%focusing on developments of the last 4 years.
\end{abstract}

%This GEO\,600 status article outlines detector developments since the
%last update article from 2010 \cite{Grote2010Status}. At that time,
%the GEO-HF program had just begun.

\section{Background}
GEO\,600~\cite{Willke2002GEO} is the German-British
laser-interferometric gravitational wave (GW) detector that was built
in the late 1990s and started operation in the early 2000s
\cite{GeoStatus2008, Grote2010GEO} in conjunction with other first
generation GW detectors. GEO\,600 uses a dual-recycled Michelson
configuration with 1200\,m long arms folded inside 600\,m long beam
tubes. From the start, the project focused on researching,
incorporating, and demonstrating riskier technologies compared to the
designs of its longer-arm counterparts, Virgo (3\,km)
\cite{Accadia2012Virgo} and LIGO (4\,km) \cite{Abbott2009LIGO}. Early
design features that were unique to GEO\,600 included multi-chain
suspensions with a monolithic final stage and a reaction mass chain,
electro-static actuators, and signal recycling
\cite{Affeldt2014Advanced}.

Following the era of first generation detector science runs when the
other observatories began going offline for substantial upgrades
\cite{Harry2010Advanced, TheVirgoCollaboration2009Advanced}, GEO\,600
opted to serve the role of being the sole detector continuing to
collect data through a program called
\emph{Astrowatch}. Simultaneously, starting in 2009, GEO\,600 began a
program called \emph{GEO-HF}~\cite{Luck2010Upgrade} which has the aim
of carrying out a set of upgrades to improve the detector's high
frequency sensitivity above approximately 600\,Hz where quantum shot
noise is the limiting noise source. The primary techniques implemented
include the injection of squeezed vacuum, a change in signal recycling
bandwidth and operating point, DC readout with an output mode cleaner,
and higher laser power combined with a thermal compensation system.

\section{GEO-HF status}
\begin{figure}[tb]
\centering
\includegraphics[width=0.8\columnwidth]{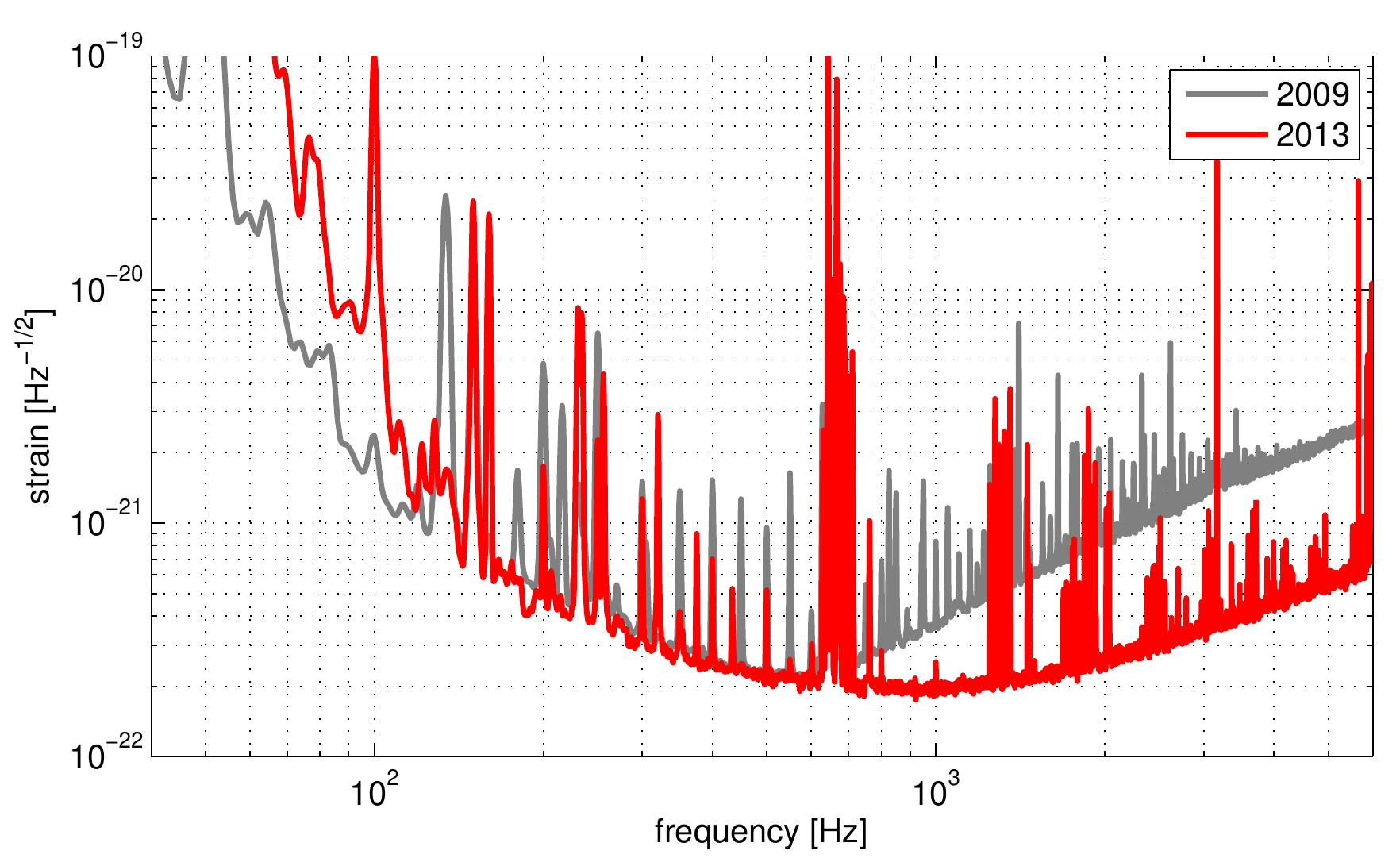}
\caption{Strain sensitivity of GEO\,600 today compared to the
  sensitivity at the start of the GEO-HF upgrade
  \cite{Grote2010GEO}. A new signal recycling mirror, squeezed vacuum
  injection, and DC readout contribute to the substantial improvement
  in high-frequency sensitivity. An effort to increase the laser power
  is ongoing. The large cluster of lines centered around 650\,Hz are
  the violin modes of the suspension fibers.}
\label{fig:h}
\end{figure}

Figure \ref{fig:h} shows the results to date of how the GEO-HF
upgrades have affected the GEO\,600 strain sensitivity. The most
notable effect is above 600\,Hz where as much as a factor of 4
improvement in sensitivity is achieved. 
%The peak sensitivity also has shifted in frequency and improved from
%$2.2 \times 10^{-22}\, \mbox{Hz}^{-1/2}$ at 600~Hz to $2.0 \times
%10^{-22}\, \mbox{Hz}^{-1/2}$ at 1000~Hz.
The dominant contributing factor to this improvement is the decrease
in the finesse of the signal recycling cavity (SRC). This was
accomplished in 2010 by exchanging the signal recycling mirror from
one with reflectivity of $R=98\%$ to one with $R=90\%$, thus reducing
the storage time of the GW signal in the interferometer and therefore
altering the detector's frequency response. The next most significant
factor is squeezing, which creates (in its current state) a factor 1.5
improvement in sensitivity above approximately 1\,kHz and was first
demonstrated in 2010 \cite{2011Gravitational}. Finally, a change in
2009 from heterodyne to homodyne (DC) readout of the GW signal
\cite{Hild2009DCreadout} renders a fundamental factor $\sqrt{3/2}$
improvement in signal-to-noise ratio (SNR)
\cite{Niebauer1991Nonstationary}. An increase in laser power is not
reflected in the 2013 sensitivity curve and is a topic of ongoing
work.
%together with a change of the SRC operation to zero de-tuning. 

The marked degradation in strain sensitivity below 100~Hz arises from
technical side effects of both DC readout and the new signal recycling
mirror. The limiting noise sources at these frequencies are alignment
feedback noise and signal recycling length feedback noise. First, the
dark-fringe offset required for DC readout creates a TEM$_{00}$
carrier field on the Michelson alignment sensors. Jitter of this field
caused by output optics creates an alignment sensing offset
which is impressed on the mirrors. Second, the new signal recycling
cavity's lower finesse decreased the SNR of the SRC length signal,
creating higher feedback noise.

%It reduces an effect called mode healing which converts higher order
%modes at the interferometer anti-symmetric port back into the
%fundamental TEM$_{00}$ Gaussian mode
%\cite{Prijatelj2012Gravitational}. The excess low frequency noise
%represents the coupling of higher order modes from suspension beam
%jitter to the GW readout.

\section{Ongoing research}
Ongoing research at GEO\,600 has several fronts. For one, there are
still some technical challenges and open research questions related to
realizing the full potential of DC readout and squeezing. Second, the
GEO-HF goal of increasing the laser power by a factor of 4\,--\,6 must
still be achieved and requires the design and commissioning of a new
thermal compensation subsystem.

\subsection{Squeezed vacuum integration}
Injection of squeezed vacuum to the interferometer's output port is a
novel technique employed to suppress quantum noise
\cite{Caves1980QuantumMechanical}. For nearly the last 3 years,
GEO\,600 has demonstrated the first long-term application of squeezed
vacuum in a gravitational wave detector, achieving a 90\% duty cycle
during science times and improving the shot-noise-limited strain
sensitivity up to a factor of 1.5 (3.7~dB) \cite{2011Gravitational,
  Grote2013First}. In order to achieve even more stable and better
squeezing levels, recent work at GEO\,600 has focused on the
development and commissioning of squeezer alignment and phase control
systems \cite{Affeldt2014Advanced}. Ongoing work includes the
commissioning of a new wavefront sensing scheme in reflection of the
OMC to ensure maximum overlap of the squeezed field with the
interferometer output field to achieve higher squeezing levels. In
addition, a new error signal for sensing the relative phase of the
squeezed field with the interferometer output is being studied in
order to eliminate lock point offsets and create more stable squeezing
\cite{Dooley2014Phase}.

%Squeezed vacuum states of light and are foreseen to be amongst the
%first upgrades to the aLIGO and aVirgo detectors.

\subsection{Higher laser power}
One unique feature of GEO\,600 is that it does not have Fabry-Perot
arm cavities. Instead, a signal recycling mirror alone is used
together with the power recycling mirror to amplify the GW signal. A
side effect is that \emph{all} of the stored power in the
interferometer is transmitted through the substrate of the beam
splitter (BS). The current operation of GEO\,600 uses 2--3~kW at the
BS, which leads to thermal lensing problems of the same magnitude that
will be faced by advanced detectors when operated at full design
power. GEO\,600 currently does not have any thermal compensation at
the BS and this limits the ability to operate with higher laser power.
During commissioning periods, stable operation with as much as 7~kW
can be achieved, but there is excess noise below 600~Hz in this high
power state. Design of an array of heating elements to project
optimized patterns of radiation onto the BS to compensate the lens
represents ongoing work.
%and the development of solutions to technical
%problems resulting from imperfect compensation represents ongoing
%work. 
Other challenges related to higher power which have already been
addressed are the correction of astigmatism at one of the end mirrors
using segmented thermal compensation \cite{Wittel2014Thermal} and the
elimination of stray light coupling to shadow sensors on each of the
mirrors through a modulation-demodulation technique
\cite{Affeldt2014Laser}.

\subsection{Alternative OMC alignment techniques}
The upgrade from RF to DC readout involved the installation of an
output mode cleaner (OMC) in the interferometer's output port directly
before the GW readout photodiode. Its purpose is to filter out all
light that does not contribute to the GW signal. Higher order modes
make it challenging to develop an alignment scheme with sufficient
bandwidth that maximizes the SNR of the light transmitted through the
OMC without introducing beam jitter coupling. Ongoing work at GEO\,600
is the commissioning of a new alignment technique which uses wavefront
sensors in reflection of the OMC. The wavefront sensors sense the
relative alignment between fields representing the GW mode and the OMC
eigenmode. The fields used are the radio frequency sidebands which are
resonant in the interferometer and audio sidebands which are generated
from a length dither of the OMC \cite{Affeldt2014Advanced}. This
scheme both eliminates low frequency dithering of the output steering
optics and facilitates an increase in the alignment control bandwidth
up to several Hz.
%, features which are helpful to GEO\,600 and which may be of interest
%for advanced detectors.

\section{Astrowatch}
Starting in 2007, GEO\,600 has served as the community's Astrowatch
detector. Astrowatch mode is a science run from the standpoint of
detector operations such that GEO\,600 is maintained at its highest
possible sensitivity in a full interferometer lock and no experiments
are allowed which could reduce its sensitivity. Furthermore, a
realtime system with interferometer parameter tracking assures valid
strain calibration \cite{Affeldt2014Advanced}. A search in the
Astrowatch data is triggered by an external event such as a gamma ray
burst \cite{Hoak2014Methods}, neutrino detection, or an optical
detection of a supernova.
%Even if unlikely given predicted event rates
%\cite{Abadie2010Predictions}, the most likely scenario for a GW
%detection by GEO\,600 would be if there were a supernova nearby in our
%Galaxy or a binary NS inspiral in our Galaxy or Andromeda.
Upon operating in science mode mostly on nights and weekends, GEO\,600
has maintained an average science time duty cycle of $62\%$ over the
last several years. Active commissioning and upgrade activities are
carried out during the remainder of the time.

\section{Summary and outlook}
GEO\,600 continues to actively pursue the goals of the GEO-HF upgrade
program and to dedicate significant time to collecting Astrowatch data
during the time when the other GW detectors are being upgraded. The
techniques of signal recycling, homodyne (DC) readout, and squeezed
vacuum injection have been demonstrated to the benefit of the GEO\,600
high frequency sensitivity. Research into better integration of
squeezed vacuum and improved OMC alignment schemes are active areas of
continued work on these topics. At the same time, a new thermal
compensation system for the BS is being developed which will enable
the use of higher laser powers. GEO\,600 will continue to collect
Astrowatch data and research advanced techniques in conjuction with
furthering the improvement of its high-frequency sensitivity.

\section*{Acknowledgements}
The authors are grateful for support from the Science and Technology
Facilities Council (STFC), the University of Glasgow in the UK, the
Bundesministerium f\"ur Bildung und Forschung (BMBF), and the state of
Lower Saxony in Germany. This work was partly supported by DFG grant
SFB/Transregio 7 Gravitational Wave Astronomy. This document has been
assigned LIGO document number P1400172.

\section*{References}
%\bibliographystyle{iopart-num}
%\bibliography{references}

\providecommand{\newblock}{}

\end{document}